\documentclass[aps,prl,twocolumn,groupedaddress]{revtex4}
\usepackage[section]{placeins}        
\usepackage{float}

\usepackage{epsfig}
\usepackage{pdfpages}
\usepackage{amsmath,amsthm, amssymb, latexsym}
\usepackage{color}
\usepackage{graphicx}
\usepackage{epsfig}
\usepackage[outercaption]{sidecap}

\newcommand{\bee}{\begin{equation}}
\newcommand{\ene}{\end{equation}}
\newcommand{\beea}{\begin{eqnarray}}
\newcommand{\enea}{\end{eqnarray}}

\begin{document}
\title{ A new mechanism of direct  coupling of laser energy to ions }
\author{Ayushi Vashistha$^{1,2,*}$, Devshree Mandal$^{1,2}$, Atul Kumar$^{1,2}$, Chandrasekhar Shukla$^3$, Amita Das$^4$}

\email{ayushivashistha@gmail.com, amitadas3@yahoo.com }

\affiliation{$^1$ Institute for Plasma Research, HBNI, Bhat, Gandhinagar - 382428, India } 
\affiliation{$^2$ {Homi Bhabha National Institute, Mumbai, 400094 } } 
\affiliation{$^3$ {Samspra Aerospace Systems Pvt. Ltd. Bengaluru, 560008} } 

\affiliation{$^4$ Physics Department, Indian Institute of Technology Delhi,  Hauz Khas, New Delhi - 110016, India }

\begin{abstract} 
The well-known schemes (e.g. Brunel, resonance absorption, JxB heating etc.) couple laser  energy to the lighter electron species of the plasma. In this work, a fundamentally new mechanism of laser energy absorption directly to the heavier ion species has been proposed. The mechanism relies on the difference between the ExB drifts of electron and ions in the oscillating electric field of the laser and an external magnetic field to create charge density perturbations. The proposed mechanism is verified with the help of Particle - In - Cell (PIC) simulations using OSIRIS4.0.

\end{abstract}
\maketitle 
\section{Introduction}
There are many well-known schemes by which the laser energy can get absorbed in the plasmas \cite{Kaw_1969, Kaw_RMP}.
For instance, resonance absorption, Brunel, JxB heating schemes etc. \cite{ping, Brunel, kruer_1985, Stix1965, Wilks1992, Freidberg}.   
All these schemes  rely on the dynamical response of the lighter electron species in the presence of the oscillating field of a high frequency laser. Consequently, the dominant transfer of energy from laser occurs in  electron species. However, there are many applications (e.g. fast ignition, medical therapy etc.) where it is  desirable to have  
  energetic  ions. This has led to  efforts seeking efficient   mechanisms of ion heating and/or acceleration \cite{ion_heating, kumar_soliton}.  
In many previous studies, a two step approach has been envisaged for this particular objective. The laser energy is first 
absorbed by electrons and its transfer to ions depends on the 
electron-ion  classical collisional and/or anomalous processes. Since Rutherford's  
 collision cross section  decreases with increasing electron energy, this process is inefficient 
 at high energies \cite{gibbon}. Many ideas of anomalous transfer of energy from electron to ion have
 also been proposed \cite{macchi, TNSA, RPA, snavely, Robinson, Fuchs2006}. In this work, we propose a novel scheme by which direct transfer of laser energy to ions is possible. 
 
The new mechanism relies on the application of a strong external magnetic field normal to the laser propagation and the polarization direction. 
 The strength of the magnetic field is chosen such that it restricts the motion of electrons which are tied to the magnetic field lines but the heavier ion species remain un-magnetized and are relatively free to move in response to the laser electric field. This happens when the electron gyroradius is much smaller than the laser wavelength whereas ion gyroradius is longer compared to it, as depicted in the schematic of Fig.1. 
  The frequency condition  $\omega_{ce} > \omega_l > \omega_{ci}$ needs to be satisfied. This requires that the applied external magnetic field strength $B_0 $ should be greater than $ m_ec\omega_l/e$.
  

 The $\vec{E} \times \vec{B}$ drift velocity in the presence of an externally applied magnetic field and an oscillating electric field is given by 
 \begin{equation}
\label{main}
 \vec{V}_{\vec{E} \times  \vec{B}}(t) = \frac{ \omega_{cs}^2}{ \omega_{cs}^2 - \omega_{l}^2} \frac{ \vec{E}(t) \times \vec{B}}{B^2}
\end{equation}
as $\omega_{ce} > \omega_l > \omega_{ci}$, the drift velocity of the two species differ  \cite{Nishikawa}. 
Here, the suffix $s = e, i $ represents the electron and ion species respectively. Thus,  $\omega_{ce}$ and 
$\omega_{ci}$ stands for the electron and ion cyclotron frequency respectively. Furthermore, $\omega_l$ 
stands for the laser  frequency. 
The difference in the drift velocities of electron and ions gives rise to a  current. 
Since, the laser electric field and the externally applied magnetic field, both are perpendicular to the propagation 
direction, this drift is along the laser propagation direction. This current has a spatial variations 
at the laser wavelength due to its dependence on the electric field of the applied laser. 
This spatial variation  leads to a finite  divergence of the current which  drives a  charge density 
fluctuation given by the continuity equation: 
\begin{equation}
\frac{\partial \rho}{\partial t} + \vec{\nabla} . \vec{J} =0
\end{equation}
This generates electrostatic plasma mode in the system, thereby, the energy of the laser  gets 
transferred to plasma. This is the underlying main principle on which our new mechanism has been proposed. It should be noted  when the applied magnetic field is zero,
 $ \vec{V}_{\vec{E} \times  \vec{B}} \approx 0$ (as $\omega_{cs}=0$) for both the species, and no charge separation will be created in the plasma.  
 In the other limit of a very large magnetic field, $ \vec{V}_{\vec{E} \times  \vec{B}} $ for both 
 ions and electrons will become same and hence in this limit also one should expect no 
 creation of number density fluctuation. Thus, in both the limits the 
 proposed  mechanism of laser energy absorption 
 will be inoperative. 
We have carried out   PIC (Particle - In - Cell ) simulations through  
OSIRIS4.0 to support our mechanism presented in this work. 

The experimental implementation of  this mechanism  requires an application of 
an external magnetic field of magnitude $B_0> m_e \omega_l c/e$. For a typical laser 
of $1\mu m$ wavelength this translates to a requirement of magnetic field of the order of 
$10^4$ Tesla. The $CO_2$ lasers which have a $10 \mu m$ wavelength, reduces the requirement by ten times.  
Lately, there has been a rapid technological progress in achieving high magnetic field in the laboratory. 
The value of 1.2 Kilo Tesla \cite{Nakamura} at Institute for Solid State Physics at the University of Tokyo, Japan has already been achieved.  Thus, it is only a matter 
of time before the requirement for carrying out experiments in this domain would be possible. 

\section{Simulation Details}

We use OSIRIS-4.0 framework \cite{hemker,Fonseca2002,osiris} for our Particle - In - Cell (PIC) 
studies. The schematics (not to scale) in Fig.1  shows the field configuration used in the simulation. The $\hat{x}$ is the propagation direction of the laser. The laser electric field is along $\hat{y}$ and the applied external magnetic field is along $\hat{z}$ as shown in the figure. 
A 2-D rectangular simulation box with dimensions  $ L_x = 3000 c/\omega_{pe}$ and  $L_y = 100 c/\omega_{pe}$ has 
been chosen. The plasma boundary starts  from   $x=  500 c/\omega_{pe} $ onwards. 
There is vacuum between $ x = 0 $ to 
 $ x = 500 c/\omega_{pe}$. 
The spatial resolution is taken  as $10$ cells per electron skin depth  corresponding to a grid size $\Delta x = 0.1 c/\omega_{pe}$ and time step for calculations is taken to be $ \Delta t = 0.02 \omega_{pe}^{-1} $.There is a sharp plasma-vacuum interface and laser is incident  from left side on the plasma target. We consider a p-polarized, plane short-pulse laser of wavelength $\lambda_l = 9.42 \mu m$ corresponding to $CO_2$ laser.  The choice of long-wavelength 
laser is simply for the sake of definiteness and is motivated by the fact that it reduces the requirement of the externally applied magnetic field. 
The mechanism, however, is independent of the laser wavelength. 
The  normally incident laser is  propagating along $ \hat{x} $, centred at  $x= 250 c/\omega_{pe}$ and 
has a pulse length ranging from   $x = 0 $  to $500 c/\omega_{pe}$. The number density of the plasma is taken to be $ n_0 = 3 \times 10^{20}$ $ cm^{-3}$ for which the  electron plasma frequency is equal to $ 10^{15} Hz$. Hence the 
 plasma is overdense for the incident laser pulse. 
 
We have followed the dynamics of both  electrons and ion species. The simulations are, however, carried out at a reduced mass  of ions which is taken to be 
$25$ times heavier than electrons ( \emph{i.e.} $m_i = 25 m_e$, where $m_i$ and $m_e$ denote the rest mass of the ion and electron species respectively). This mass ratio has been chosen 
to reduce the computational time. (The computational resources being meagre at our disposal.) 
  We use normalized units henceforth, unless otherwise stated explicitly. 
The time is normalized by the inverse of electron plasma frequency corresponding to the chosen 
plasma density of $n_0$. The laser frequency is $\omega_l = 0.2 \omega_{pe}$.
The length is normalized by electron skin depth $c/\omega_{pe}$ and  the 
magnetic field by $m_ec \omega_{pe}/e$. The external magnetic field has been chosen in such a way that 
$ \omega_{ce} > \omega_{l} > \omega_{ci}$.
To satisfy this condition the value 
of magnetic field is chosen  to be $B_0 = 2.5$. For this field, the electrons gyrate at a frequency  
$ 2.5 \omega_{pe} $.  The number of particles per cell are taken to be $4$. Boundary conditions for fields and particle are periodic in transverse direction and absorbing in longitudinal direction. The laser is taken to be infinite in y-direction.

The simulation geometry is specifically chosen to avoid the  possibility of any well-known absorption schemes 
to be operative. The laser frequency being $0.2$ is much smaller than the electron plasma frequency (unity here). 
The laser is  incident normal to  the sharp plasma interface. This ensures the absence of  
 resonance and vacuum heating schemes. Also, the role of $\vec{J} \times \vec{B} $ electron heating  
 is   made negligible 
 by choosing the laser intensity to lie  in the non-relativistic domain of $a_0 = e E/m_e \omega_l c < 1$. 
 Thus the  peak intensity of laser in simulation is taken to be  $I = 3.5\times 10^{15} W/cm^2$ with a rise and fall time of $205 \omega_{pe}^{-1}$  each. 
%
%
\section{Observation}
 We first provide a comparison between two cases, namely (A) for which the applied magnetic 
field $B_0 = 0$ and (B) for which $B_0 = 2.5$. The choice of $B_0 = 2.5$ ensures that the condition 
$\omega_{ce} > \omega_l > \omega_{ci}$ is satisfied. The simulation configuration being symmetric in the $y$ coordinate, we choose to depict the fields as a function of 
the coordinate $x$ in  the figures. 

 In Fig \ref{fig2}, the three subplots in the first row depict  the plot of the time-dependent magnetic field   associated with the  laser and the self-consistent response of the plasma medium as a function of $x$ at three distinct time 
 (i) $t = 0$ showing the initial laser pulse field when it does not touch the plasma surface, (ii) $t = 350$ when the laser 
 has touched the plasma and (iii) $t = 700$ when the laser pulse has been reflected from the plasma surface 
 and the plasma is left with the remnant self-consistent excitations. The dashed red line denotes the 
 case (A) for which $B_0 = 0$ and the solid blue line corresponds to case (B) with $B_ 0 = 2.5$. 
  It can be observed from the figure that in  case(A) there is no penetration of the laser magnetic field in the plasma 
  and almost a  complete reflection of the pulse occurs. 
  However,  for  case (B) when the external applied field $B_0$ is finite and equal to $2.5$, a part of the laser 
  field shows clear penetration in the plasma medium.  In the same figure 
  the next row of three subplots show the plasma current $J_x$ as a function of $x$ for the two cases. 
  The third and the last row depicts the electrostatic field $E_x$ generated in the plasma medium. 
  It is clear from the plots of $J_x$  and $E_x$ that while the plasma gets stirred significantly  by the laser field 
  in the presence of applied external magnetic field in case(B), for case(A) the plasma  continues to remain quiescent even after 
  interacting  with the laser pulse. This is as expected by  our proposed mechanism.

 We now compare the evolution of various energies for the two cases in Fig \ref{fig3}.  We indicate distinct  time incidents by vertical lines in this figure. 
 The first two solid black vertical lines denote the times $t_1 = 100$  and 
 $t_2 = 380$, where $t_2-t_1$ region indicates the time interval during which the laser interacts with the plasma. 
 Thereafter,  laser field gets reflected from the plasma surface. At around $t_3 = 500$ denoted by the black dashed line, 
 the laser reflects from the plasma surface and at $t_4 = 780$ indicated by black  dotted  line the laser field leaves the 
 simulation box.  The evolution of the electromagnetic field energy has been shown by the blue dashed and solid lines 
 for case(A) and case(B) respectively. The left side of the $y$-axis in the figure represents the scale for this 
 particular energy. The electromagnetic field energy remains constant associated with the laser pulse till $t_1$. 
 At $t_1$ the laser pulse interacts with the plasma surface. There is a fall of electromagnetic energy from $t_1$
 to $t_2$, the time for which the laser pulse interacts with the plasma medium.  It is evident from the figure 
 that the electromagnetic energy decreases more in case(B) than in case(A) during this period.  From 
 $t_2$ to $t_3$ when the laser pulse reflects from the plasma surface but remains within the 
 simulation box the electromagnetic energy remains constant.  At $t_3$ the left edge of the reflected laser pulse 
 touches the left boundary of the simulation box. Thereafter there is a continuous decay in the field 
 energy till $t_4$ at which the pulse completely leaves the simulation box. Thus, while  the  first fall of the electromagnetic field energy between $t_1$ to $t_2$ is associated with the laser pulse interacting with the plasma medium, the  second fall 
 of the electromagnetic energy between $t_3$ to $t_4$ is due to the laser pulse leaving the simulation box. 
 The laser energy transfer, therefore, essentially occurs between $t_1$ to $t_2$. Further, it should be noted that the energy transfer to plasma medium is more in case(B) with externally applied magnetic field.  
 Clearly, this energy transfer to plasma would be in the form of field and kinetic energies. 
 Another interesting thing to observe is that even after the laser pulse leaves the system, the electromagnetic 
 energy for case(B) remains finite in contrast to case(A) where it falls off to zero. This shows 
 that the laser interaction creates some electromagnetic field fluctuation in the plasma medium in case(B).

  The evolution of the kinetic energy of electrons and ions are shown by green and red color line plots 
  respectively in the 
  same Fig \ref{fig3}. The dashed and dotted lines are for case(A) and the  solid line with dots correspond 
  to case(B). For these  kinetic energy plots, the right $y$-axis defines the scale.  
  In the absence of applied magnetic field in case(A) the gain in ion energy is much smaller compared to
   that of electrons. However, for case (B) in the presence of magnetic field the kinetic energy of ions 
   is much higher than that of  electrons. The total energy absorbed  by plasma in this case is higher. 
%
%
%
 It is clear that  the laser interacts with  plasma in case (B) in a more 
 efficient fashion and   excites certain disturbances in the plasma.  The disturbances excited in case(B) 
 do have an electrostatic 
 characteristic as evidenced from the evolution of the $x$ component of the electric field $E_x$ 
 shown in Fig.\ref{fig4}(a) at a location $x = 540 c/\omega_{pe}$ (which is deep in the bulk 
 region of  the plasma medium). For case(A) the amplitude of the electrostatic field $E_x$ is negligible. 
 The frequency associated with the   regular oscillation in the electric field $E_x$ is identified by carrying a 
 Fourier transformation which is shown in 
   Fig \ref{fig4}(b).  The peak in the Fourier transform for the two cases appear at very distinct frequencies. 
    In case(A) the spectrum shows a peak at 
  $\omega=1$ which is the electron plasma frequency. On the other hand, for  Case(B) the peak in the frequency is  shifted at a lower side (corresponds to a value close to  $\sim 0.2$) which indicates towards presence of an ion dominated mode.

    The  oscillations in the electrostatic field are a result of  space charge fluctuations 
     excited in the plasma. The number density and the velocity  fluctuations associated with the two species 
     have been shown in the first and second row of the figure for the two species of electron and ions by  
     blue and red color lines respectively as a function of $x$ at three different times (Fig \ref{fig5}). The third row of subplots shows the current density fluctuations.  
     It is clear from the plot that except for the short distance near the plasma boundary at $x= 500$,  the ion and electron density oscillations and their $x$ component of the velocities are in tandem. 
     These oscillations propagate 
     deeper in the plasma with time. The laser interacting at the plasma boundary excites  a space 
     charge fluctuations at the boundary  which then travels in  the  bulk region of the plasma. 

In Fig.6 we show a plot of the velocity and density fluctuations of the two species as a function of time 
for $x = 540$ (a point in the bulk region of  the plasma). The plots correspond to 
 six different values of the magnetic field. From the plots it is clear that at low values of the magnetic field 
the velocity  fluctuations are small. At an intermediate value of the magnetic field $B_0 =2, 2.5, 3.0$ 
the velocity fluctuations increase and the difference between electron and ion velocity is also high. 
The number density fluctuations are also high for these cases. However, at very high value of magnetic field 
of $B = 10$ the electron and ion velocities are almost similar and the number density fluctuation is small. 
It should be noted that at the intermediate value of $B= 2, 2.5, 3.0$ only the required condition of 
$\omega_{ce} > \omega_l > \omega_{ci}$ is satisfied. For $B=10$  even the ions get  magnetized. This is consistent with our proposed mechanism in which the number density fluctuation are supposed to be excited by the difference in the $\vec{E} \times \vec{B}$ velocity of the two species. The difference in the drift velocity of the two species vanishes in both the limits of low and high magnetic fields. 
  

      These number density fluctuations  associated with ion time scale of response 
   peak at an intermediate value of $B_0$ the applied magnetic field. The mode disappears both at the 
   low magnetic field and also at the higher values of the magnetic field. To understand the role of magnetic field 
   on laser absorption, we show 
   in Table I the total energy absorbed by the two species electrons and ions from the laser for various values of the 
  applied magnetic field. From this table, it is clear that the energy absorption in ions is low 
  both at the low value of magnetic field and also at the high value of the magnetic field. 
  This has also been depicted in  the  plot of  Fig \ref{fig7}  showing the total energy absorption by electrons and ions at various values of the external magnetic field. We have also shown total current $J_x$ in the system.  The ion absorption follows the profile of the current as a function of the applied magnetic field. 
  
  The particle count of the two 
  species as a function of energy has been shown in Fig \ref{fig8} for the two cases of (A) and (B).  From this figure also 
  one can infer clearly that energy absorption by ions  in the presence of external magnetic field is higher. 
  
%


%
 \section{Conclusion}

In conclusion,  the results of the previous section clearly demonstrate 
 that the laser energy gets preferentially coupled to ions in the presence of 
  external magnetic field. The space charge fluctuations created by the 
  difference in the $\vec{E} \times \vec{B}$ velocity of the two species in presence of oscillating 
  electric and static external magnetic field appears responsible for the excitation of the electrostatic mode in the 
  plasma medium.

  While we have  primarily focused here on the illustration of a fundamentally new concept, it should be realised 
  that in recent years there has been rapid progress in the creation of high magnetic fields in the laboratory. 
  Magnetic field of the order of $1.2 $ kilo Tesla has already been produced in laboratory \cite{Nakamura, Yoneda_2012}.  One expects that 
  in near future this limit will increase. We feel that it is only a matter of time when we will witness magnetic 
  field of the order of $10s$ of kilo Tesla at which the mechanism presented in this manuscript can be experimentally verified in laboratory with a long wavelength $CO_2$ lasers which have now already  been made to operate in the pulsed high-intensity mode \cite{Haberberger2010, Tochitsky2016, Beg1997, Fujioka2016}.

\paragraph{}
\paragraph{}

\paragraph{\bf{{Acknowledgment:}}}

\paragraph{}
\paragraph{}

The authors would like to acknowledge the OSIRIS Consortium, consisting of UCLA ans IST(Lisbon, Portugal) for providing access to the OSIRIS4.0 framework which is the  work supported by NSF ACI-1339893. AD would like to acknowledge her  J. C. Bose fellowship grant JCB/2017/000055 and the CRG/2018/000624 grant 
of DST for the work. The simulations for the work described in this paper were performed on Uday, an IPR Linux cluster.
AV, DM, AK and CS would like to thank Dr. R.K. Bera for fruitful discussions.

\bibliographystyle{ieeetr}  

\bibliography{atul_thesis_new}

\paragraph{}
\paragraph{}

\begin{figure*}
	
	\includegraphics[width=0.5\linewidth]{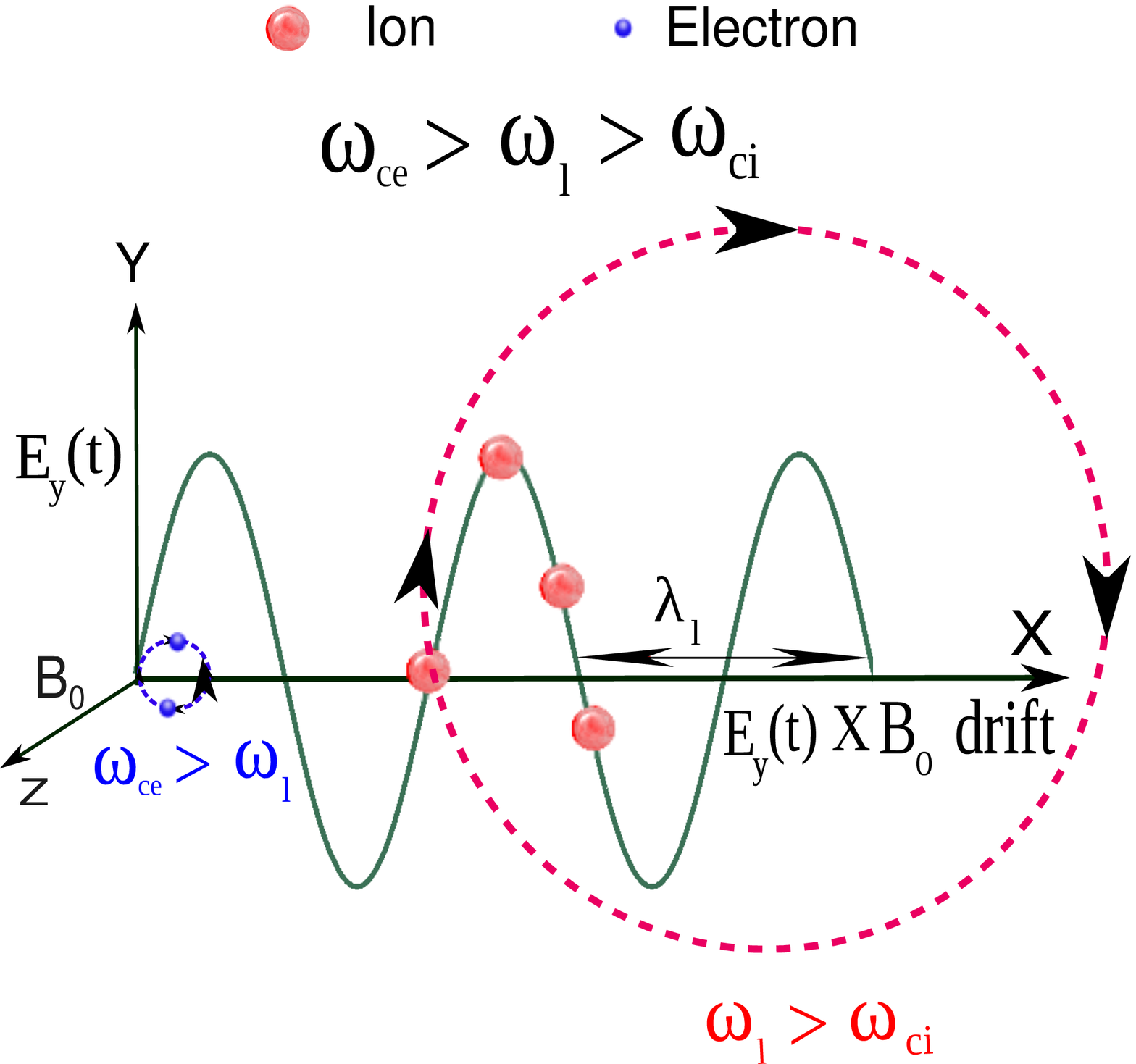}
	
	\caption{Schematic (not to scale) showing that under the condition $ \omega_{ce} > \omega_{l} > \omega_{ci}$, the electrons are tightly bound to the external magnetic field (along $  \hat{z} $ ) whereas ions are free to move under the effect of laser. The direction of oscillating electric field due to laser is along $ \hat{y} $ and that of laser magnetic field along with a static external magnetic field is along $ \hat{z}$ and the laser propagation is along $ \hat{x} $ direction. $\vec{E} \times \vec{B} $ drift due to oscillating electric field is acting along $ \hat{x} $ }
	
	\label{fig1}
\end{figure*}%

  \begin{figure*}

  	\includegraphics[width=1.0\linewidth]{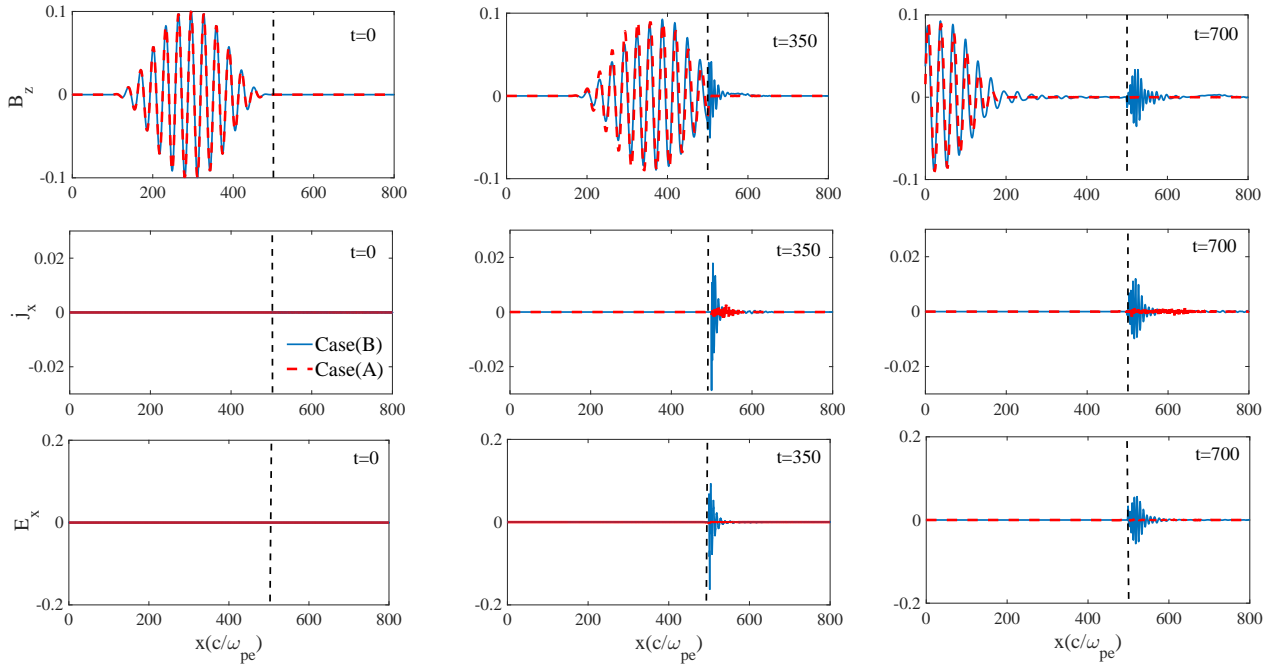}
  	\caption{Time evolution of spatial variation of ${B_z}$, x-component of total current $J_{x}$ and x-component of electric field $E_{x}$ along $\hat{x}$, averaged over $\hat{y} $,  showing that as laser interacts with plasma(t=350), there is a finite magnitude of perturbed $B_z$,$J_{x}$ and $E_{x}$ in case(B) which is absent in case(A). From these plots, we can infer that a part of laser energy has been coupled into plasma. }
  		
  	\label{fig2}
  \end{figure*}%
  
  \begin{figure*}

  	\includegraphics[width=1.0\linewidth]{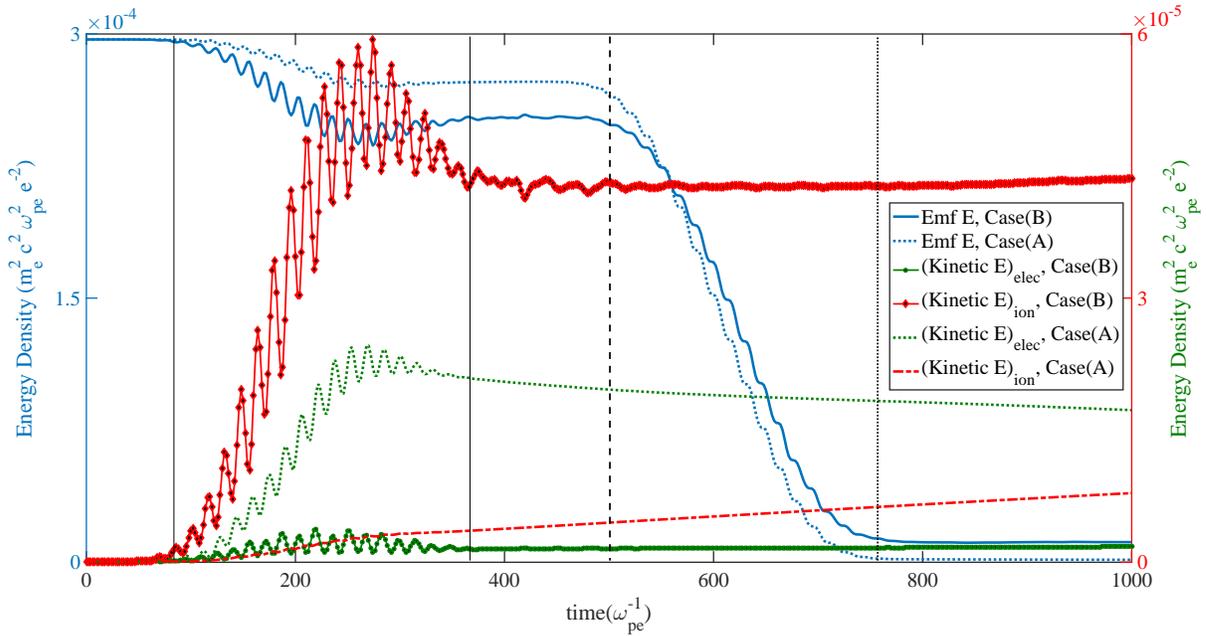}
  	\caption{Time variation of mean of Electromagnetic field (Emf) energy density of the system (left y axis) and kinetic energy density of both the species (right y axis) showing that laser energy is mainly gained by electrons in case(A) and by ions in case (B). The decrease in Emf energy of the system between the two solid lines is due to the interaction of laser with plasma (there is gain in Kinetic energy of electrons in case (A) and ions in case(B) at the same time). Between dashed and dotted line, the laser reflects back and moves out of the simulation box and as a result, we observe a drop in Emf energy of the system. There is also generation of some electrostatic mode in plasma in case (B) contributing to emf energy of the system even after the laser has been reflected out of the simulation box (beyond dotted line). }
  	\label{fig3}
  \end{figure*}%

  \begin{figure*}
  
  	\includegraphics[width=1.0\linewidth]{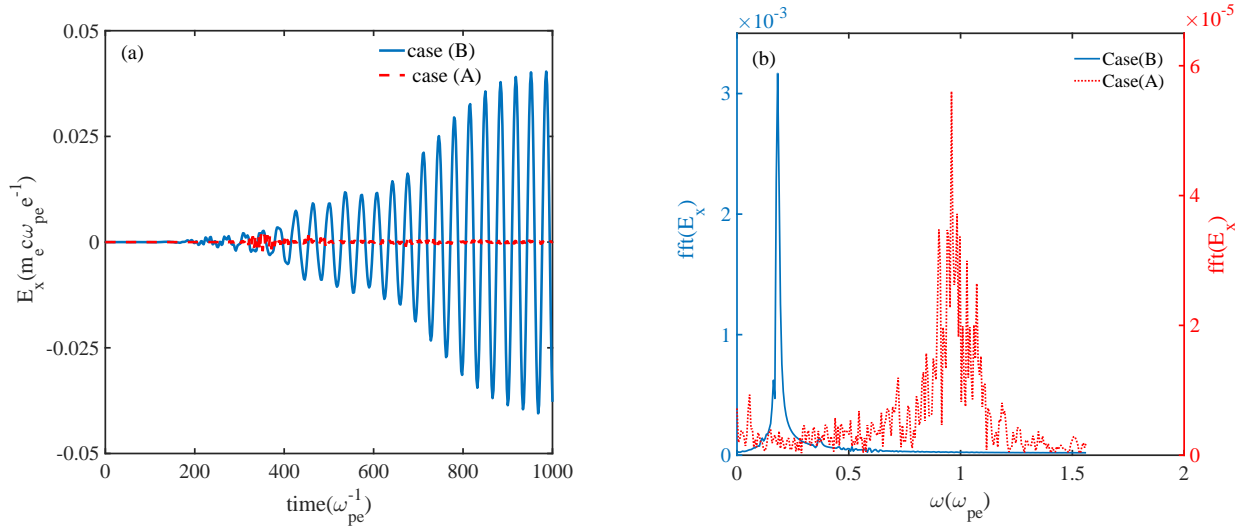}
  	
  	\caption{ \textbf{(a)}  Time variation of x-component of electric field $ E_x $ for both the cases at x=540 $c/\omega_{pe}$, averaged over $\hat{y} $, showing generation of electrostatic mode in the bulk plasma in case (A) and case(B). \textbf{(b)}  Fast fourier transform of $E_x$ showing peak at electron plasma oscillation for case (A) and at a lower frequency for case (B). The shift in the frequency towards lower end in case (B) indicates presence of an ion dominated mode.}
  	\label{fig4}
  \end{figure*}%

  \begin{figure*}
 
  	\includegraphics[width=1.0\linewidth]{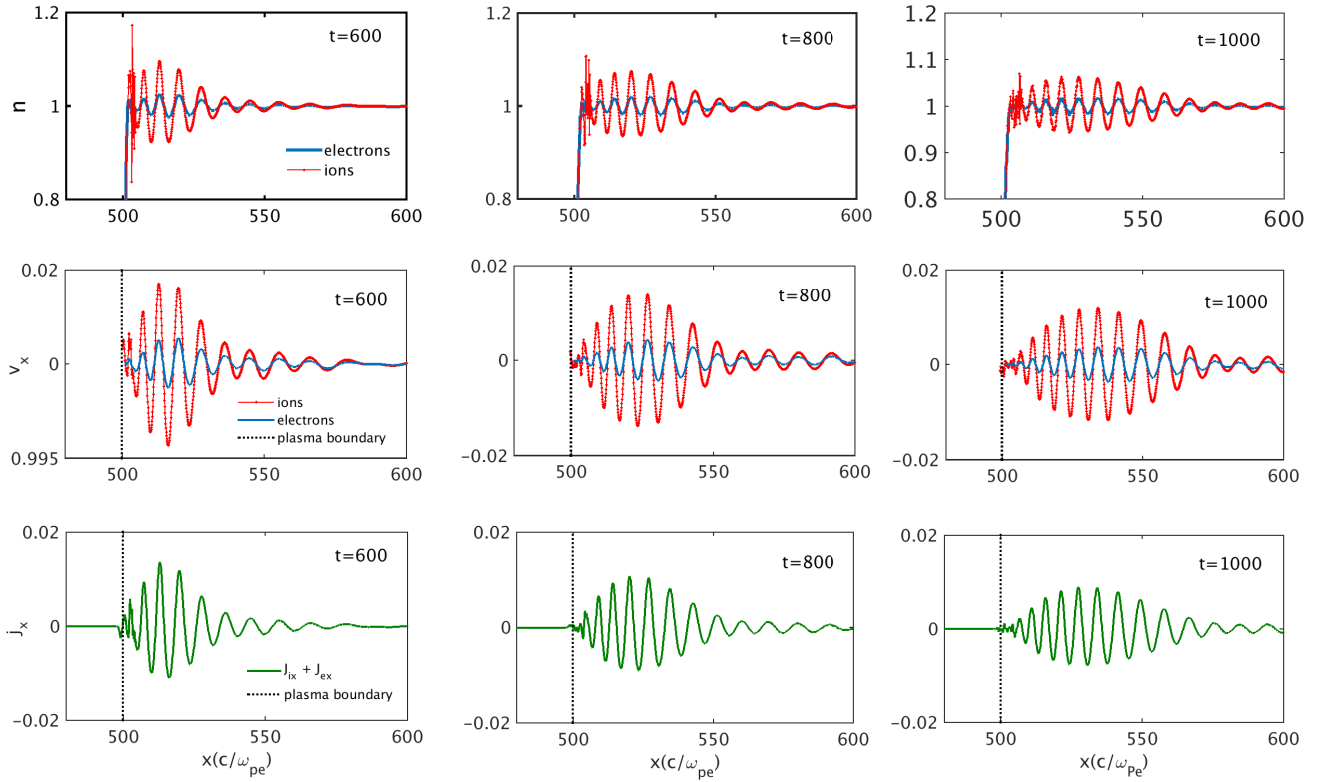}
  	\caption{ Figure shows the spatial variation of density (n), x-component of velocity $V_{x} $ and x-component of total current $J_x$ averaged over $\hat{y} $ at different times. At a particular time we observe that the density perturbations created and the difference in $V_{ix} - V_{ex} $ are at the same position, indicating the correspondence between them. }
  	\label{fig5}
  \end{figure*}

  \begin{figure*}
  	
  	\includegraphics[width=1.0\linewidth]{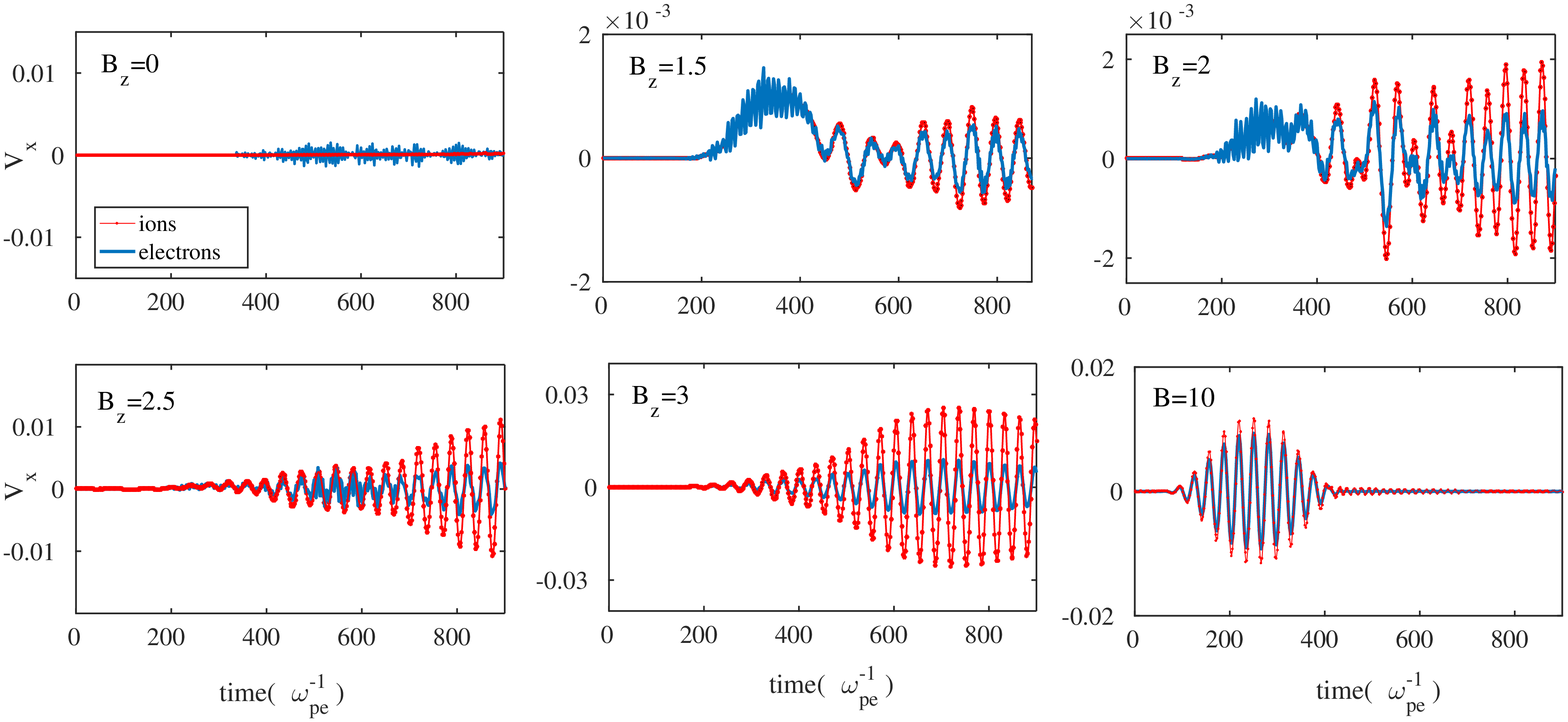} 
  	\includegraphics[width=1.0\linewidth]{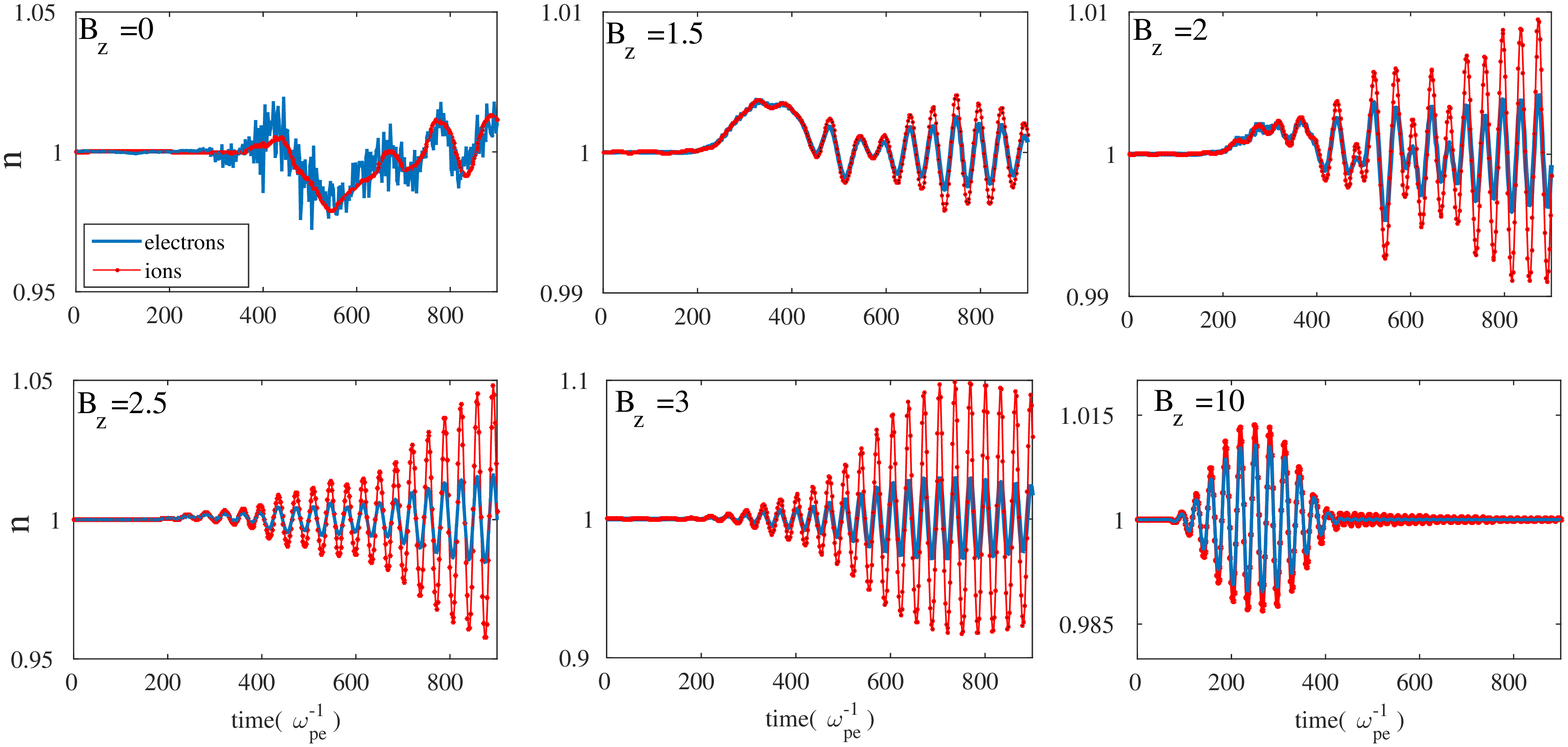} 
  	\caption{ Figure showing the effect of varying external magnetic field on $ {V_{x}}(t) $ and density perturbations of both the charged species at a particular x (here,x=540), averaged over $\hat{y} $. With increasing  applied magnetic field, $V_{ix} - V_{ex}$, when $\omega_{ce} > \omega_{l} > \omega_{ci}$ is satisfied . As soon as $\omega_{ce} > \omega_{l} > \omega_{ci}$ is not satisfied ($B_{z}=10$), $V_{ix} - V_{ex}$ again decreases. Similar trend is observed in density perturbations of both the species as well. }
  	\label{fig6}
  \end{figure*}%

 \begin{table*}
 	\begin{tabular}{|p{3cm}||p{3.0cm}||p{3.0cm}||p{3cm}|}
 		\hline
 		&\% of laser energy absorbed by electrons&\% of laser energy absorbed by ions & \% of laser energy absorbed by plasma\\
 		\hline
 		\hline
 		$B_{z}=0$&6.06&2.40&8.46\\
 		\hline
 		$B_{z}=1.5$&0.42&2.76&3.18\\
 		\hline
 		$B_{z}=2$&0.42&9.68&10.10\\
 		\hline
 		$B_{z}=2.5$&0.58&14.52&15.10\\
 		\hline
 		$B_{z}=3$&1.22&31.12&32.34\\
 		\hline
 		$B_{z}=3.5$&1.41&44.60&46.01\\
 		\hline
 		$B_{z}=4$&0.22&10.98&11.2\\
 		\hline
 		$B_{z}=5$&0.003&0.61&0.61\\
 		\hline
 		$B_{z}=7$&0.002&0.002&0.004\\
 		\hline			
 		$B_{z}=10$&0.002&0.005&0.007\\
 		\hline	
 		
 	\end{tabular}\caption{Table showing that laser energy absorbed by ions increases as applied external magnetic field increases. It can also be noted that the ion energy is significant for the case only when  $V_{ix} - V_{ex}$  is finite (Fig \ref{fig6}) which is only true if  $\omega_{ce} > \omega_{l} > \omega_{ci}$ is satisfied.}.
 \end{table*}

   \begin{figure*}

  		\includegraphics[width=0.7\linewidth]{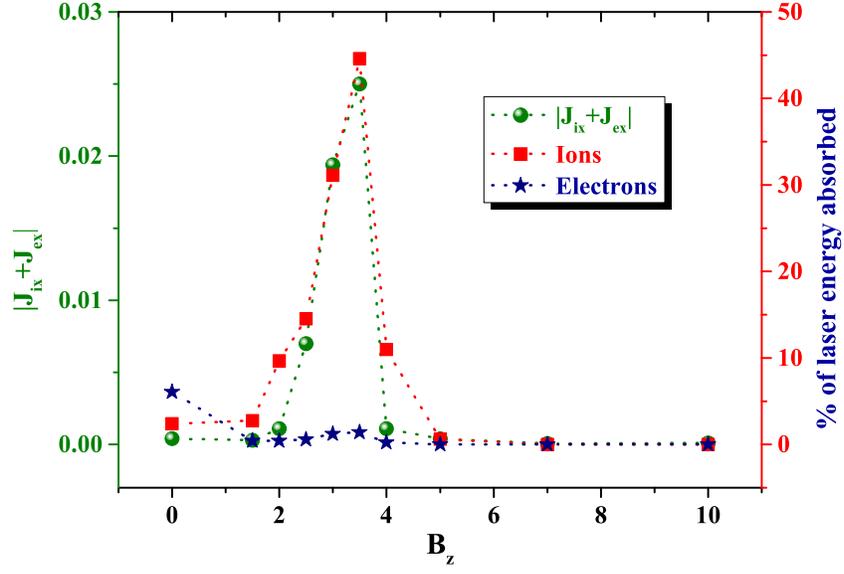} 
   	\caption{Figure shows the effect of varying external magnetic field on the total current generated in the system (left y axis) and energy absorbed by electrons and ions (right y axis). It can be observed from the graph that the energy absorbed by ions follow the total current in the system. Also, total current as well as energy absorbed by ions increase only when $ \omega_{ce} > \omega_{l} > \omega_{ci}$. }
   	\label{fig7}
   \end{figure*}%

      \begin{figure*}
     
      	\includegraphics[width=1.0\linewidth]{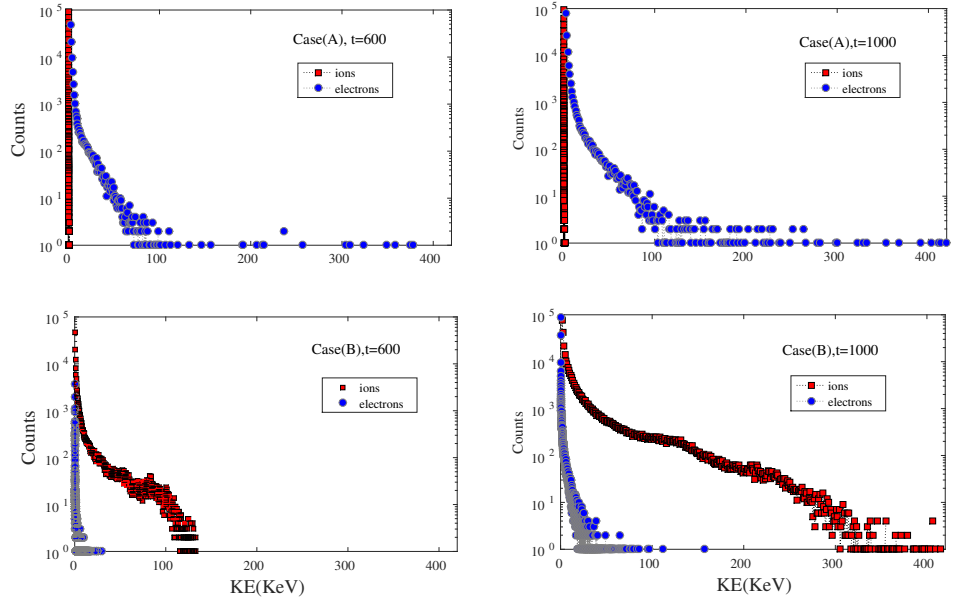}
      	\caption{Energy distribution of ions and electrons for case(A) and (B) showing the maximum energy gained by them in both the cases. }
      	\label{fig8}
      \end{figure*}
   
  \end{document}